\documentclass[a4paper,12pt]{article}
\usepackage{amsmath}
\usepackage{epsfig,amssymb}
\title{Variational analysis of self-focusing of intense ultrashort pulses 
in gases}
\author{ E. Ar\'evalo and A. Becker\\
{\it Max-Planck-Institut f\"ur Physik komplexer Systeme}\\
{\it  N\"othnitzer Strasse 38, D-01187 Dresden, Germany}
}
\date{\today}

\begin{document}

\maketitle

\begin{abstract}
By using perturbation theory we derive an expression for the electrical
field of a Gaussian laser pulse propagating in a gas medium. This expression is used as
a trial solution in a variational method to get quasianalytical
solutions for the width, intensity and self-focusing distance of
ultrashort pulse. The approximation gives an improved agreement with
results of numerical simulations for a broad
range of values of the input power of the pulse than previous analytical
results available in the literature.
\end{abstract}

\section{Introduction}

The propagation of high-peak power femtosecond laser pulses in optical media
has attracted considerable attention recently \cite{Kasparian03,Alfano89}. These pulses undergo dramatic
changes in their temporal, spatial and spectral properties due to their
nonlinear interaction with the medium. The most fundamental process  
is self-focusing which causes the pulse to be compressed in space, resulting 
in an increase of the peak intensity \cite{marburger75}. Self-focusing would result in a 
catastrophic
collapse of the pulse, but the self-focusing process is usually balanced
by multiphoton ionization or excitation of the atoms or molecules in
the medium, since the resulting (quasi-)free electron density defocuses the 
pulse. 

The analysis of optical pulse propagation is usually based on the description 
of the pulse
in terms of its complex field envelope, neglecting the underlying
rapid oscillations at its carrier frequency. The resulting slowly varying 
envelope approximation reduces the Maxwell's equations to higher dimensional 
nonlinear Sch\"odinger equations (NLSE).
These equations are not integrable, so they do not 
have soliton solutions. However, they possess stationary solutions which are  
unstable on propagation. The self-focusing distance is determined as the
(first) point of infinite intensity in the solution of the NLSE.

Most of the quantitative analysis of self-focusing (and further
propagation) of the pulse results from numerical computation. For example, from a
curve-fitting procedure based on numerical simulations a popular 
analytical formula for the location of the first singularity 
(self-focusing distance) for the propagation of a CW laser
beam has been given by Dawes and Marburger 
\cite{dawes69,marburger75}:
\begin{equation}\label{zsfmar}
\xi^{(M)}_{sf}=\frac{0.367}{\sqrt{(\sqrt{\cal P}-0.852)^2-0.0219}},
\end{equation}
where $\cal P$ is the input power scaled in units of the critical power for
self-focusing, $P_{crit} = \lambda_0^2/2\pi n_0n_2$, with $\lambda_0$ is the 
wavelength and $n_0$ and $n_2$ are the linear and nonlinear
indices of refraction, respectively \cite{fibich00}.
Formula (\ref{zsfmar}) has been verified in many experiments and
numerical simulations. Besides its accurateness and usefulness for estimations
it however does not give much insight in the physics of the 
underlying process. 

In order to get a deeper insight into a physical process, it is often useful
to use approximative theories, even if full numerical
solutions are available. A few approximations have been discussed
already in the
early review paper on self-focusing by Marburger \cite{marburger75}. The
resulting expressions provide, however, rather qualitative than quantitative 
estimations only. 
Alternative approaches have been proposed to
analyze the effect of self-focusing, namely a systematic perturbation theory
\cite{fibich96,fibich99}, a ray-equation approximation
\cite{marburger75}, 
an approach based on Fermat's principle
\cite{boyd},  a variational analysis \cite{akozbek00}, a paraxial parabolic 
approximation \cite{Schwarz-Diels01}, or a source-dependent expansion
method \cite{Sprangle}. 
The predictions of these models for the self-focusing distance agree 
quantitatively well with the exact solutions from numerical simulations 
for input powers close to the critical power, but deviate quickly as the power
increases.

In this paper we present a quasianalytical approximation which provides a 
description of the self-focusing phenomenon not only qualitatively but also
quantitatively. It is based on the variational approach using a trial
solution, which contains a first-order perturbation correction of the phase 
resulting from the transverse distortion of the pulse. As we show below, 
predictions for the self-focusing distance, the on-axis peak intensity 
and the radius of the pulse within this ansatz are in better agreement 
with those of numerical simulations over a broad range of input
powers than previous analytical results available in the literature. We show
how the pulse separates into an inner and an outer 
self-trapped component along the propagation distance. 
The inner component self-focus, while the outer one stays
as background.

The paper is organized as follows: First, we derive the phase correction
of the pulse using first-order perturbation theory. The corrected form of the 
pulse is then used as trial solution in the Lagrangian of the system to
obtain equations for the self-focusing distance, the on-axis peak intensity
and the radius of the pulse. Next, the predictions of the approximation will be
compared with those of the earlier models and of numerical simulations.
Finally, an analogy with the problem of a particle in a
finite quantum well is presented. In this analogy, the bound state and its
exponential-decaying wings penetrating a smooth potential wall
correspond to  the inner and outer component of the pulse,
respectively. The self-focusing is associated with the shrinking of
the width of the quantum well.

\section{Variational approach  \label{refsec1}}

We study the propagation of a linearly polarized laser beam in gases.
For the explicit calculations we do not restrict to any specific gas. 
Our analysis is based on the
scalar wave equation, which can be obtained from the Maxwell equations (e.g.
\cite{boyd}):
\begin{equation}
\label{max1}
\partial_z^2E+\Delta\,E-\frac{\epsilon_0}{c^2}\partial_t^2E=
\frac{4\pi\chi^{(3)}}{c^2}\partial_t^2E^3
+\frac{4\pi}{c^2}\partial_tJ.
\end{equation}
Here, $\chi^{(3)}$ is the third order nonlinear 
susceptibility coefficient of the 
medium, $\epsilon_0$ is the linear dielectric constant of the gas and c is the
speed of light. 
Since we are interested in the dynamics of the pulse up to the self-focusing
distance only, we neglect the plasma effects ($J=0$), which usually balance the
catastrophic collapse and, hence, influence mainly the propagation of the
pulse after the self-focusing point. Using the slowly varying envelope approximation, 
Eq. (\ref{max1}) can be written in the retarded coordinate frame 
$(t\rightarrow t-z/v_g)$ as:
\begin{eqnarray}\label{sch2}
&&i\,\partial_{\xi}\,u+\frac{1}{4}\Delta_{\perp}\,u+
\frac{L_D}{L_{nl}}\vert u\vert^2\,u 
\nonumber\\
&&
-\frac{1}{4}\frac{L_D}{L_d}
\bigg(\partial_{\tau}^2u+\frac{i}{3}\frac{L_d}{L_d^{\prime}}\partial^3_{\tau}u\bigg)=0,
\end{eqnarray}
where $u={\cal E}/\sqrt{I_0}$, $\xi=z/L_D$, $\tau=t/T_0$, and
the transverse coordinate $r$ is given in units of the length of the
pulse $w_0$ ($w_0$ is the radius at $1/e^2$ of the irradiance). The characteristic lengths
are given by 
$L_D = k_zw_0^2/2$, 
$L_d = T_0^2/2k_2$,
$L_{d^{\prime}} = T_0^3/2k_3$ and 
$L_{nl} = (n_2k_0I_0)^{-1}$, where
$k_0=2\pi/\lambda$, $k_z=n_0k_0$.
$k_2$ and $k_3$ are the second and third order group-velocity dispersion 
coefficients, respectively, $T_0$ is the 
duration of the pulse ($T_{FWHM}=\sqrt{2\,\log{2}}\,T_0$), and $I_0$  is the input peak
intensity (in units of W/cm$^2$). 
We further neglect in Eq. (\ref{sch2}) the effect of the group velocity 
dispersion to get the well-known (2+1)-dimensional NLSE,
\begin{equation}\label{nlse1}
i\,\partial_{\xi}\,u+\frac{1}{4}\Delta_{\perp}\,u+
F[u\,u^{*}]\,u =0,
\end{equation}
where the star stands for complex conjugation and 
$F[u\,u^{*}]=c_1\vert u\vert^2$
with
$c_1=L_D/L_{nl}$.
During the early stage of propagation of
the pulse (up to the self-focusing point) in gases 
the effect of the group velocity is negligible, namely if
$L_{nl} \ll L_d < L_{d^{\prime}}$.
Thus, for the present analysis we expect that the condition is fulfilled
as long as the self-focusing distances are small, i.e.\ as larger the input
power of the beam is.

\subsection{Phase correction}

\begin{figure}
\centerline{\epsfxsize=8.0truecm \epsffile{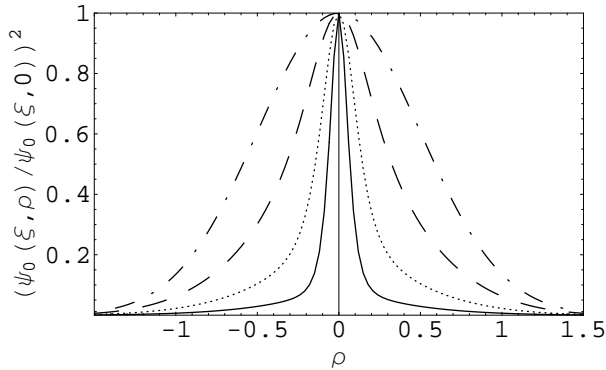}}
\caption{Results of numerical simulations for the
normalized pulse shape for $P/P_{cr}=5$: $\xi=0$ (dotted-dashed line),
$\xi=0.15$ (dashed line), $\xi=0.20$ (dotted line), $\xi=0.23$ (solid line).}
\label{shapes}
\end{figure}

The variational method has been applied to the (2+1)-dimensional
NLSE recently \cite{akozbek00} to study qualitatively the phenomenon
of self-focusing in air. As mentioned at the outset the
method is as good as the trial solution. 
Usually the self-similarity assumption is used in order 
to integrate over the transversal coordinates taking
into account that the total power of the NLSE is invariant.
This is done by assuming that the shape of the pulse remains 
unchanged up to the self-focusing point. The assumption preserves 
the lens transformation of the NLSE. However, from numerical simulations 
one can observe that the initial pulse shape gets distorted during
the self-focusing process (c.f.\ Fig.\ \ref{shapes}). In fact, 
the pulse separates into  two components as it propagates 
\cite{fibich99},
\begin{equation}\label{pertur1}
u(\xi,r)=\psi_0(\xi,r)+\epsilon\psi_1(\xi,r),
\end{equation}
where $\psi_0$ is the high intensity inner core of the pulse, which
self-focuses and $\epsilon\psi_1$ is the low intensity outer part, 
which propagates forward following the linear propagation mode. 
In recent experiments and numerical simulations 
\cite{Skupin04,Mlejnek98,Kandidov03,Liu} it has been shown 
that this weak large background plays a decisive role in the propagation and
filamentation process. 

We expect that the self-focusing process is influenced by the interaction 
between the inner core and the outer part too. To this end we use a trial
solution in the variational method, in which the perturbation
by the background is taken into account as a correction in the phase. 
We first separate amplitude and phase of the pulse as:
\begin{eqnarray}
u(\xi,r)&=&\vert u(\xi,r)\vert \exp{(i\,S(\xi,r))}\nonumber\\
&=&
\vert\psi_0(\xi,r)\vert \exp{(i\,S_0(\xi,r))}
\nonumber\\
&&+
\epsilon\vert \psi_1(\xi,r)\vert\exp{(i\,S_1(\xi,r))},
\end{eqnarray}
and by substituting in Eq. (\ref{nlse1}) we get  up to order $\epsilon$ that
\begin{equation}\label{phase1}
S(\xi,r)=\sum\limits_{j=0}^{1}S_j(\xi,r)+ O(\epsilon^2)
\end{equation}
with
\begin{equation}\label{finalphase}
S_j(\xi,r)=S_j(0,r)+\int\limits_{0}^{\xi}n_j(\xi,r)d\xi,\quad j=0,1.
\end{equation}
Here,
\begin{eqnarray}\label{index0}
n_0(\xi,r)
&=&
F[\vert\psi_0(\xi,r)\vert^2]-\frac{1}{4}
(\partial_{\xi}S(\xi,r))^2
\nonumber\\
&&
+\frac{\Delta_{\perp}\vert\psi_0(\xi,r)\vert}{\vert\psi_0(\xi,r)\vert}
\end{eqnarray}
and
\begin{eqnarray}\label{index1}
&n_1(\xi,r)&=
\epsilon\bigg\{\bigg(
2F^{\prime}[\vert\psi_0(\xi,r)\vert^2]\vert\psi_0(\xi,r)\vert
\vert\psi_1(\xi,r)\vert
\nonumber\\
&-&\frac{\vert\psi_1(\xi,r)\vert\Delta_{\perp}\vert\psi_0(\xi,r)\vert}{4\vert\psi_0(\xi,r)\vert^2}+
\frac{\Delta_{\perp}\vert\psi_1(\xi,r)\vert}{4\vert\psi_0(\xi,r)\vert}
\nonumber\\
&-&\frac{\vert\psi_1(\xi,r)\vert}{4\vert\psi_0(\xi,r)\vert} (\partial_{r}S(\xi,r))^2   
\bigg)
\nonumber\\
&\times & \cos{(S_0(\xi,r)- S_1(\xi,r))}
\nonumber\\
&+&\bigg(\frac{\vert\psi_1(\xi,r)\vert\Delta_{\perp}S_1(\xi,r)}{4\vert\psi_0(\xi,r)\vert}
\nonumber\\
&+&\frac{\partial_{r}\psi_1(\xi,r)\partial_{r}S_1(\xi,r)}{2\vert\psi_0(\xi,r)\vert}
\bigg)
\nonumber\\
&\times & \sin{(S_0(\xi,r)- S_1(\xi,r))}\bigg\}.
\end{eqnarray}
The desired trial solution is then given by
\begin{equation}\label{ansatz1}
u(\xi,r)=\psi_0(\xi,r)\exp{(i\,S_1(\xi,r))},
\end{equation}
with $S_1(0,r)=0$. Here the amplitude corrections of order
$\epsilon$ have been neglected, while phase corrections have been kept. 
Note that $u(\xi,r)$ still fulfils the self-similarity
assumption, namely 
$\vert u(\xi,r)\vert^2=\vert \psi_0(\xi,r)\vert^2$.

The phase correction $S_1(\xi,r)$ is obtained using first-order perturbation
theory as follows. 
Inserting Eq. (\ref{pertur1}) into Eq. (\ref{nlse1}) and collecting powers
of $\epsilon$ we get
\begin{eqnarray}\label{pertur3a}
\epsilon^0 &:&
   i\,\partial_{\xi}\,\psi_0+\frac{1}{4}\Delta_{\perp}\,\psi_0+F[\psi_0\psi_0^{*}]\,\psi_0=0,\\
\label{pertur3b}
\epsilon^1 &:& 
   i\,\partial_{\xi}\,\psi_1+\frac{1}{4}\Delta_{\perp}\,\psi_1+F[\psi_0\psi_0^{*}]\,\psi_1\nonumber\\
&&+F^{\prime}[\psi_0\psi_0^{*}]\psi_0\psi_0^{*}\psi_1+F^{\prime}[\psi_0\psi_0^{*}]\psi_0^2\psi_1=0,
\end{eqnarray}
where the prime stands for the derivative with respect to the argument. 
Note, that the solution of Eq. (\ref{pertur3b}) has the form
\begin{equation}\label{psi1}
\psi_1=\partial_{a(\xi)}\psi_0,
\end{equation}
where $a(\xi)$ is the length of the pulse. Thus, the phase correction can be
obtained for {\it any shape} of the inner core by taking its derivative
and separating the phase.

We do not consider the presence of any
external perturbation, like losses or ionization. However, the theory
presented here can be extended considering these terms following the
same lines as in \cite{akozbek00}. 
We have checked carefully that the predictions of the numerical
calculations for the self-focusing distance do vary by less 
than $1\%$, if we include external perturbations. Other properties, such as
the pulse length or the on-axis intensity (up to the self-focusing point) are
even less sensitive to these effects. It is therefore justified, for the sake
of simplicity, to neglect external perturbations.

\subsection{The Gaussian pulse shape}

As outlined above, the phase correction can be obtained for any shape of the
pulse. In order to investigate the effect we apply the theory to the most
important case of a Gaussian pulse shape below. In this case
\begin{equation}\label{psi0gauss}
\psi_0(\xi,r)=\frac{A}{a(\xi)}
\exp{(\rho^2)}
\exp{(i\,b_0\,a(\xi)\,\rho^2)},
\end{equation}
where $\rho=r/a(\xi)$ and $b_0$ is a constant describing
the initial wave front divergence of the pulse (i.e. $b_0=0$
for collimated pulse, or $b_0=-1/f$ for the case of an external 
lens with focal length $f$), we obtain the phase correction 
from the solution of Eq. (\ref{pertur3b}),
\begin{eqnarray}\label{psi1gauss}
\psi_1(\xi,r)&=&\frac{A}{a^2(\xi)}\exp{(-\rho^2)}
\exp{(i\,b_0\,a(\xi)\,\rho^2)}
\nonumber\\
&&
\times \bigg(2\rho^2-i\,b_0,a(\xi)\,\rho^2-1
\bigg),
\end{eqnarray}
as follows.
Substituting Eqs. (\ref{psi0gauss}) and (\ref{psi1gauss}) into
(\ref{index0}) and (\ref{index1}) we get
\begin{eqnarray}\label{index1gauss}
&n_0(\xi,r)&=-\frac{1}{4}(\partial_{r}S(\xi,r))^2
\nonumber\\
&+&\frac{1}{a^2(\xi)}\left(\rho^2-1+A^2c_1\exp{(-2\rho^2)}\right),\nonumber\\
\end{eqnarray}
and
\begin{eqnarray}\label{index2gauss}
&n_1(\xi,r)&=\frac{2\,A^2\,c_1\,\epsilon}{a^3(\xi)}
\left(2\rho^2\exp{(-2\rho^2)}-\exp{(-2\rho^2)}\right)
\nonumber\\
&+&\frac{2\epsilon}{a^3(\xi)}\left(1-2\rho^2\right),
\end{eqnarray}
where $a(\xi)$ is an undetermined function.
In Ref. \cite{fibich96} it has been shown that 
$a(\xi)$ can be determined using
\begin{equation}\label{eqbeta}
\beta=-a^3(\xi)a^{\prime\prime}(\xi),
\end{equation}
which is proportional to the excess power, as long as the
excess power is small. With this relation 
one is able to solve the integral (\ref{finalphase}) in
terms of $\beta$ for Gaussian pulses. 
However, the solution of the integral is ill-posed, since the 
integrals in the solution diverge. 
Therefore, we have adopted an adiabatic
approximation instead, namely that the
transverse form of $n_1(\xi,r)$ remains unaffected along the
propagation distance, i.e
\begin{eqnarray}\label{finals1}
S_1(\xi,r)&=&n_1(\xi,r)\xi\\
\label{finals2}
         &=&\frac{2\,A^2\,c_1\,\epsilon\,\xi}{a^3(\xi)}
\left(
2\rho^2\exp{(-2\rho^2)}-\exp{(-2\rho^2)}\right)
\nonumber\\
&+&\frac{2\epsilon\xi}{a^3(\xi)}\left(1-2\rho^2\right).
\end{eqnarray}
The last term of Eq. (\ref{finals2}) arises due to  the
diffraction term in Eq. (\ref{pertur3b}) for the outer part of the pulse.  
This term should not effect the propagation of the inner core part and
we neglect it. The validity of this assumption can be shown using the
variational method (see Appendix \ref{app2}). 
Thus, finally we get 
\begin{eqnarray}\label{ansatz2}
&u(\xi,r)
&=
\frac{1}{a(\xi)}A\exp{\left(\rho^2\right)}
\nonumber\\
&\times&
\exp{\left(i\,b_0\,a(\xi)\,\rho^2\right)}
\exp{(iS_1(\xi,r))}
\end{eqnarray}
with
\begin{eqnarray}
S_1(\xi,r)=\frac{2\,A^2\,c_1\,\epsilon\,\xi}{a^3(\xi)}
(2\rho^2-1)\exp{(-2\rho^2)}.
\end{eqnarray}

\subsection{Lagrangian and variational analysis}

The Lagrangian functional for the system (\ref{nlse1})
reads,
\begin{equation}\label{lagrangian}
{\cal L}=\frac{i}{2}(u\,\partial_{\xi}u^{*}- u^{*}\partial_{\xi}u)
+\frac{1}{4}\vert\partial_{\xi}u\vert^2-\frac{1}{2}c_1\vert u\vert^2.
\end{equation}
From Eqs. (\ref{ansatz2}) and (\ref{diffracrestric}) we define 
\begin{eqnarray}\label{ansatz3}
u(\xi,r)&=&\frac{1}{a(\xi)}A(\xi)
\exp{\left(-\rho^2\right)}\exp{\left(i\,b_0\,a(\xi)\,\rho^2\right)}
\nonumber\\
&&\times
\exp{\left(i\phi(\xi)\right)}\exp{(iS_1(\xi,r))}
\end{eqnarray}
with
\begin{eqnarray}\label{ansatz3a}
S_1(\xi,r)=c(\xi)a(\xi)
\left(2\rho^2-1\right)\exp{(-2\rho^2)}, 
\end{eqnarray}
where the variational parameter $A(\xi)$, $a(\xi)$, $c(\xi)$ and
$\phi(\xi)$ are sufficient to describe the dynamics of the problem.
Here, we consider a collimated input laser pulse, focused by a lens, given by
\begin{equation}\label{var10}
u(0,r)=
u_0\,\exp{(-\tau^2)}\exp{(-r^2)}\exp{\bigg(-i\frac{r^2}{f}\bigg)}.
\end{equation}
where $f$ is the focal length of the lens in units of diffraction
length $L_D$. Note that the temporal dependence of the variational
parameters in (\ref{ansatz3}) is not written explicitly.
Inserting Eq. (\ref{ansatz3}) into the Lagrangian
(\ref{lagrangian}) and integrating over the transverse coordinate we obtain
the reduced Lagrangian 
\begin{equation}\label{var11}
\langle {\cal L} \rangle=2\pi\int\limits_{0}^{\infty} {\cal L}\,r\,dr,
\end{equation}
which depends on the variational parameters and the independent
variables $\xi$ and $\tau$ only. 
The equations of motion for the variational parameters are given by
\begin{equation}\label{var12}
\frac{d}{d\xi}\partial_{(\partial_{\xi}\mu_i)} \langle {\cal L}
\rangle-\partial_{\mu_i} \langle {\cal L} \rangle=0,
\end{equation}
where $\mu_i$= $A$, $a$, $c$, $\phi$ (i=1,...,4). This leads to
the following set of coupled equations:
\begin{eqnarray}\label{var13}
a^{\prime}(\xi)=\frac{16}{9}c(\xi)+b_0,
\end{eqnarray}
\begin{eqnarray}\label{var14}
c^{\prime}(\xi)=-\frac{2({\cal P}(\xi)-1)}{a^3(\xi)},
\end{eqnarray}
\begin{eqnarray}\label{var15}
{\cal P}^{\prime}(\xi)=0,
\end{eqnarray}
where ${\cal P}(\xi)=P(\xi)/P_{cr}$ with
$P(\xi)=2\pi\int\limits^{\infty}_{0}\vert u\vert\, r\, dr =\pi A^2({\xi})/2$,
and the critical power $P_{cr}=\pi/c_1$.

From Eqs. (\ref{var13})-(\ref{var15}) with the initial conditions,
$a(0)=1$ and $c(0)=0$,
we get 
\begin{equation}\label{eqtosolve}
a^3(\xi)a^{\prime\prime}(\xi)=-\beta=-\frac{32}{9}({\cal P}-1).
\end{equation}
or 
\begin{eqnarray}\label{width1}
a(\xi)&=&\sqrt{(1-\,b_0\,\xi)^2-\beta\xi^2}.\\
\end{eqnarray}
The self-focusing distance, 
\begin{eqnarray}
\label{selffocus1}
\xi_{sf}&=&\frac{1}{\sqrt{\beta}-b_0},
\end{eqnarray}
is obtained for the 
collapse of the pulse ($a(\xi)=0$). Note that
\begin{equation}
\frac{1}{\xi_{sf}}=\frac{1}{\xi_{sf}(b_0=0)}+\frac{1}{f},
\end{equation}
in agreement with the lens transformation property of the NLSE \cite{marburger75,fibich99}.
The on-axis intensity is given by
\begin{eqnarray}\label{effectiveinten}
I(\xi)&=&\vert u(\xi,0)\vert^2
=\frac{A^2(\xi)}{a^2(\xi)}\nonumber\\
&=&\frac{A^2(\xi)}{(1-\,b_0\,\xi)^2-\beta\xi^2}.
\end{eqnarray}

\section{Results and comparisons}

Below we present the predictions of the present theoretical ansatz and
compare them with those of other semi-analytical estimations and the results
of numerical calculations using Eq. (\ref{nlse1}). The comparisons are performed for the case
of a collimated Gaussian pulse without external focusing. We should
note that this is the most extreme case. The presence of a finite external
focusing reduces the relative errors between the theory and simulations.

\begin{figure}
\centerline{\epsfxsize=8.0truecm \epsffile{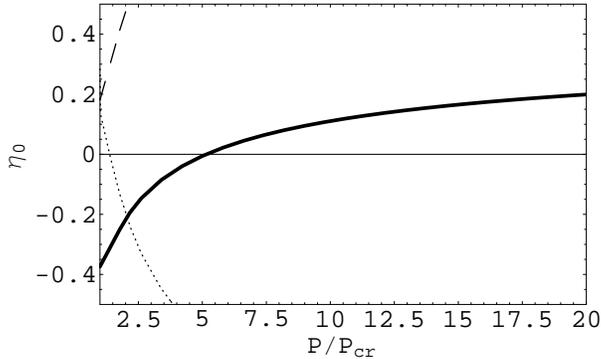}}
\caption{Relative error $\eta_{sf}$ of the predictions of the present
theory for the self-focusing distance (solid line) with respect to the
Marburger's Formula as function of ${\cal P}=P/P_{cr}$. Also shown are
the relative errors for earlier models (see text).}
\label{figrelerror0}
\end{figure}

First, we concentrate on the results of the present ansatz for 
the self-focusing distance,
$\xi_{sf}$ (c.f. Eq. (\ref{selffocus1})). In Fig. \ref{figrelerror0} 
the relative error (solid line),
\begin{equation}\label{relerror}
\eta_{sf} = \frac{\xi_{sf}-\xi_{sf}^{(M)}}{\xi_{sf}^{(M)}},
\end{equation} 
with respect to Marburger's formula, Eq. (\ref{zsfmar}), is
shown as a function of the input power (scaled in units of the critical
power). Note that Eq. (\ref{zsfmar}) has been derived from numerical
simulations and is, hence, an excellent estimate of the exact self-focusing distance.
The comparison shows that the relative error is largest (about
$40\%$) for input powers close to the critical power and is  
less than $20\%$ for $2.5<{\cal P}<20$. The large error near the critical
power for self-focusing might be due to the neglect of the group velocity in
the analysis. 

Also shown in Fig. \ref{figrelerror0} are the relative errors (with respect
to Marburger's formula) for the predictions of the self-focusing distance
resulting from analyses 
by means of either one of the following methods, 
ray-equation approximation \cite{marburger75}, Fermat's principle
\cite{boyd}, variational analysis \cite{akozbek00}, paraxial parabolic 
approximation \cite{Schwarz-Diels01}, or source-dependent expansion
method \cite{Sprangle} (dashed line):
\begin{equation}\label{zsf2}
\xi_{sf}^{(1)}=\frac{1}{\sqrt{{\cal P}-1}},
\end{equation}
and by using perturbation theory (dotted line, \cite{fibich99}): 
\begin{equation}\label{zsf1}
\xi^{(2)}_{sf}\sim \frac{2}{{\cal P}}\sqrt{\frac{M}{N_c\sqrt{{\cal P}-1}}},
\end{equation}
where the constants $M=0.55$ and $N_c=1.86$ are derived from the Townes
soliton shape. 
The value of $\xi^{(2)}_{sf}$ in Eq. (\ref{zsf1}) is given in
diffraction units and therefore four times larger than in \cite{fibich99}. 
The latter formula, Eq. (\ref{zsf1}), is valid for ${\cal P}<2$.

Both estimations are in excellent agreement with the results of Marburger's
formula near the critical power but strongly deviate as the power increases
at larger powers. 
It is clearly seen, that already at ${\cal P} \simeq 2$ the predictions within the
present theory are in closer quantitative agreement with the exact
results. For example, at ${\cal P}=10$ the predictions of
the earlier theories deviate from the exact result by more than 
$70\%$ in the case of perturbation theory (see Eq. (\ref{zsf1}))
or by more than $100\%$ in the
case of other theories (see Eq. (\ref{zsf2})), while the present
theoretical prediction agree within about $20\%$ with Marburger's formula.
We attribute this to the phase correction obtained from the
interaction with the weak
large background, which has not been taken into account in the earlier
work. In fact, from our ansatz we observe that the background does
not diffract and receives energy from the inner part of the pulse
during the self-focusing process, as we will discuss below in Section \ref{quantum}.
We also mention that the expression in Eq. (\ref{zsf2}) for the
self-focusing can be obtained by taking into account the two first leading
terms of the power series of the phase correction in
Eq. (\ref{ansatz3a}) only. This means that the earlier ansatz (see for
instance \cite{akozbek00}) is a limiting case of 
Eqs. (\ref{ansatz3}) and (\ref{ansatz3a}).

\begin{figure}
\centerline{\epsfxsize=8.0truecm \epsffile{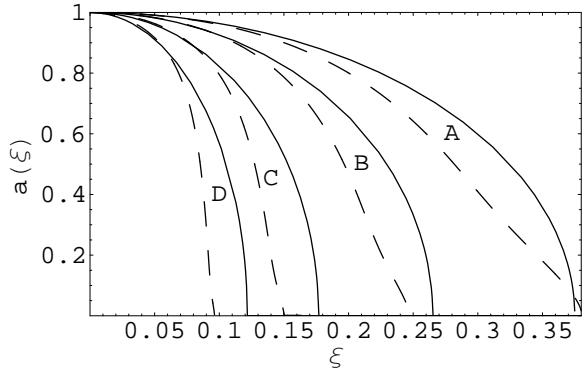}}
\caption{Comparison of the predictions of the present theory (solid
lines) for the pulse length with numerical simulations (dashed lines)
for ${\cal P}=3$ (A), ${\cal P}=5$ (B),
${\cal P}=10$ (C), ${\cal P}=20$ (D).} 
\label{width}
\end{figure}

We find also that the estimations of the present
theory and the numerical simulations for the pulse length and the
on-axis intensity agree within 30\%. 
In Fig. \ref{width} a comparison of the pulse length $a(\xi)$,
Eq. (\ref{width1}) (solid line), and the $1/e^2$-level width of
the intensity profile using numerical simulation (dashed line) of a collimated
Gaussian pulse for different input powers is shown. 

\begin{figure}
\centerline{\epsfxsize=8.0truecm \epsffile{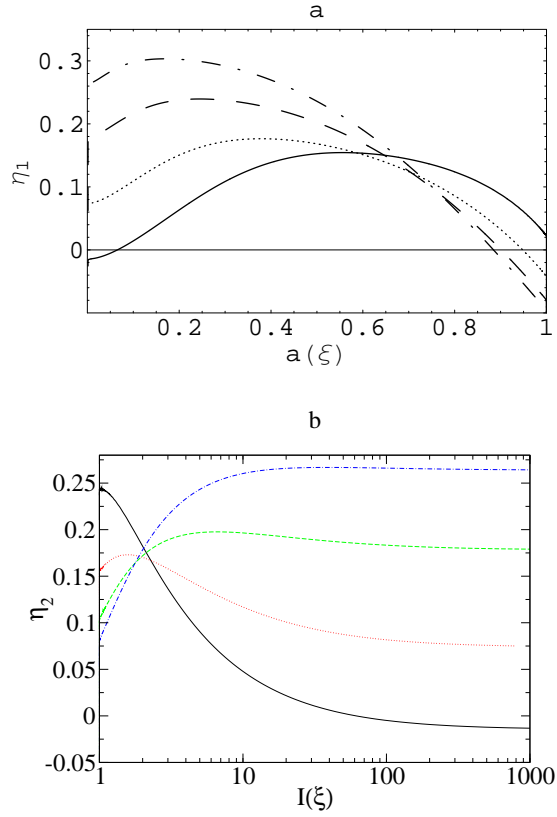}}
\vspace{0.5cm}
\centerline{\epsfxsize=7.0truecm \epsffile{relerr2.eps}}
\caption{Relative errors $\eta_1$ (panel a) and $\eta_2$ (panel b) for
different input powers, ${\cal P}=3$ (solid line),
${\cal P}=5$ (dotted line), ${\cal P}=10$ (dashed line), ${\cal P}=20$
(dotted-dashed line).} 
\label{figrelerror1}
\end{figure}

Since the range of
values of the ordinate in Fig. \ref{width} is the same (but not those of the abscissa) 
we have evaluated 
an relative error using the inverse functions, 
\begin{equation}\label{relerror1formula}
\eta_1=\frac{\xi[a_{theory}]-\xi[a_{simulation}]}{\xi[a_{simulation}]}.
\end{equation}
The results are presented
in Fig. \ref{figrelerror1}(a) as a function of $a(\xi)$ for several
values of the input power. We observe that $\eta_1$ is in general
lower than $10\%$ for $a(\xi)\sim 1$, i.e. in the early stages
of the propagation (c.f Fig. \ref{width}). Close to the self-focusing
point, $a(\xi)\sim 0$, $\eta_1$ increases to above $20\%$ for high input
powers (${\cal P}\geq 10$). This shows that in the simulations 
the pulse undergoes a stronger deformation than predicted by the present
ansatz. 

In the same way we define
the relative error for the on-axis intensity, namely
\begin{equation}\label{relerror2formula}
\eta_2=\frac{\xi[I_{theory}]-\xi[I_{simulation}]}{\xi[I_{simulation}]}.
\end{equation}
As can be seen from the results in
Fig. \ref{figrelerror1}(b) $\eta_2$ is lower than $20\%$
for ${\cal P}\leq 10$.
In general, $\eta_2$ tends to be constant for larger
values of the intensity.

\section{Quantum well analogy \label{quantum}}

\begin{figure}
\centerline{\epsfxsize=7.0truecm \epsffile{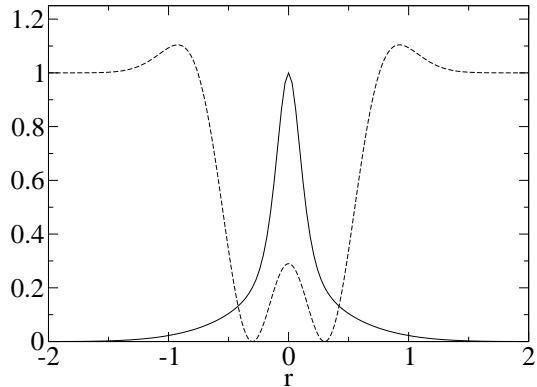}}
\caption{Comparison between the intensity profile  of the pulse (solid
line) obtained from the simulation and the trapping potential derived
from the theory at $\xi=0.2$ with ${\cal P}=5$.}
\label{figpotpulseprofile}
\end{figure}

In order to further understand
the role of the weak large background
on the dynamics of the pulse, we have investigated our 
ansatz using the quantum well analogy. 
In this picture $\xi$ acts as a fictitious time variable,
$u(\xi,r)$ as a bound state and $-\phi^{\prime}(\xi)$ as the energy
function of the system. 
By substituting Eq. (\ref{ansatz3}) into Eq. (\ref{nlse1}) and taking
the real part, one finds  an 
equation for $-\phi^{\prime}(\xi)$. If we neglect the diffraction part
of this equation (neglecting kinetic energy in the quantum
analogue), it is straight forward to derive a potential-like
function. In Fig.  \ref{figpotpulseprofile} the normalized shape of
this potential is compared with the normalized shape of the intensity
profile  of the pulse obtained from numerical simulations.
We observe that the shape of the potential is a well which contains
two regions, an inner and an outer part.  
The inner part extends from the center of the well  ($r=0$)
up to the absolute minimum.
The outer part includes the rest, namely the potential wall which
smoothly grows radially up to a certain constant value.  Notice that the
wall is infinitely thick, so no analogue of the tunnel effect is
possible here. These features of the potential well 
reveal already  that the pulse possesses two components. In fact, we
observe in Fig. \ref{figpotpulseprofile}
that the inner component of the pulse is bell-shaped, while the outer
component is an analogue of the exponential-decaying wings of the
probability density of a bound state penetrating a smooth potential well.
The potential well shrinks radially along the
propagation distance and the local maximum in the center
decreases and eventually disappears at the self-focusing
distance. So, the inner component of the pulse tends to concentrate at the
origin and eventually self-focuses. 

\begin{figure}
\vspace{1cm}
\centerline{\epsfxsize=7.0truecm \epsffile{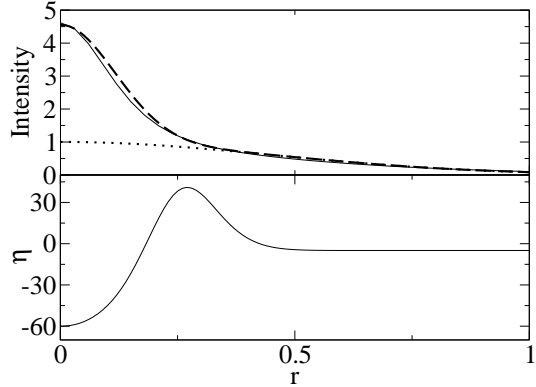}}
\caption{The upper panel contains an example of the intensity
distribution obtained from simulation (solid line) for ${\cal P}=5$
and $\xi=0.2$. The distribution can be fitted by the superposition of two
Gaussians (dashed line). One for the inner part, and the other for
the background (dotted line). The lower panel contains the result of the
operation in Eq. (\ref{operation}). Note that the region, where the
dashed line tends towards the dotted line (inner and outer
part join), is located where the
maximum of the function $\eta$ is.}
\label{operation1}
\end{figure}

We have estimated the region in which the inner part of the pulse is
concentrated as follows.
From the intensity profile of the pulse, obtained from numerical simulations,
one can notice that it is well fitted  by the superposition of two Gaussian
functions, as shown in the upper panel of Fig. \ref{operation1}. 
In order to determine quantitatively where the inner and outer part join, we
have performed the following operation in our simulations:
\begin{equation}\label{operation}
\eta=\partial_r\bigg(\frac{\partial_r\,\vert u(\xi,r)\vert^2}{\vert u(\xi,r)\vert^2}\bigg),
\end{equation}
where $\vert u(\xi,r)\vert^2$ is the intensity distribution.
Note that $\eta$ is constant when $u(\xi,r)$ is a pure Gaussian
function. 
For the present case (c.f. upper panel of
Fig. \ref{operation1})  $\eta$ has its absolute maximum in the region
where the inner and outer part of the pulse join (c.f. lower panel of
Fig. \ref{operation1}). Thus, we determine the
boundary position, $r_{join}$, between the inner and
outer part of the pulse by the absolute maximum of $\eta$.

\begin{figure}
\centerline{\epsfxsize=8.0truecm \epsffile{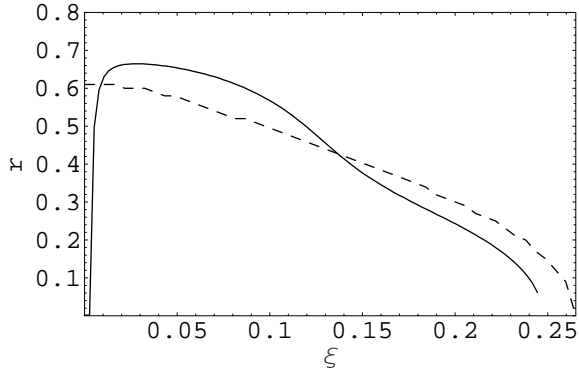}}
\caption{Comparison of the position of the minimum of the potential (dashed line) and
the position of the deformation of the Gaussian pulse  (solid line)
vs. $\xi$ for ${\cal P}=5$.}
\label{minradius}
\end{figure}

\begin{figure}
\centerline{\epsfxsize=8.0truecm \epsffile{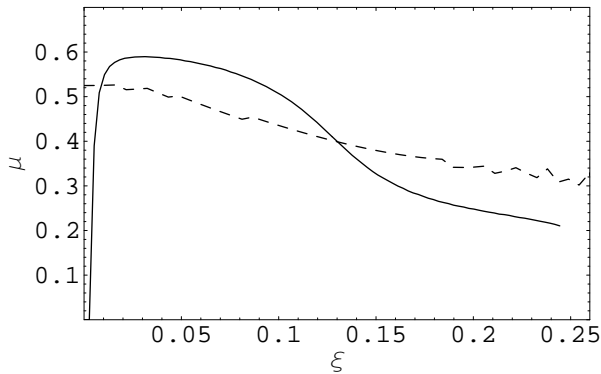}}
\caption{Estimated fraction of energy, $\mu$, that self-focuses for 
${\cal P}=5$. Solid line: simulation, dashed line: theory.}
\label{minener}
\end{figure}

In Fig. \ref{minradius} we show the comparison
between  $r_{join}$ from simulations (solid line)  and the position of the absolute minimum of the
potential obtained from the quantum well potential (dashed line c.f. Fig. \ref{figpotpulseprofile})
Note that at $\xi=0$ no deformation is observed in the simulations since
the initial pulse shape is Gaussian. For $\xi >0$ a deformation of the
pulse appears at the origin ($r=0$) and moves quickly to $r\simeq
0.65$. Afterwards $r_{join}$ shrinks and disappears at the
self-focusing distance. This qualitative behavior is found for all the cases
considered in this paper. 
Thus, 
our theoretical model does not predict the appearance of the
deformation at the origin, but the position of the absolute minimum of the
potential well follows closely the behavior of $r_{join}$ over the main
part of the propagation distance.

From the radii shown in
Fig.\ref{minradius} we can calculate the fraction of energy $\mu$ stored in
the inner component of the pulse, which is compared in Fig.
\ref{minener} with those from the numerical simulations. 
We observe that there is a  
energy flow from the inner part to the outer part when the pulse
shrinks. Note that the inner component of the pulse releases more
energy in the second half of its propagation towards the self-focusing 
distance, where the shrinkage of $r_{join}$ is more pronounced. 
At the self-focusing distance most
of the energy is stored in the outer component of the pulse which
stays with the pulse and does not diffract.
We also note that 
for low powers (${\cal P}<5$) the present model overestimates the
release of energy towards the outer region, while
for higher powers  (${\cal P}>5$) it is underestimated. This
difference is related to the
discrepancy found for the estimation of the self-focusing distance
(see Fig. \ref{figrelerror0}). We note, that the ansatz
in earlier works (see for instance \cite{akozbek00}) would lead to
a potential well too, however the wall of this potential well has a finite
thickness. Thus, in the previous ansatz the outer part of the pulse
diffracts out in contrast to the present model where the pulse is
completely self-trapped.

\section{Conclusion}

In conclusion, we have revisited the problem of self-focusing of a
laser pulse in a gas Kerr medium modeled by the 
(2+1)-dimensional nonlinear Schr\"odinger equation. We have defined a
new trial solution taking into account that during the self-focusing
process the pulse splits in an inner and an outer component. In the trial
solution the outer component is taken into account via a 
phase correction. We have used a quantum well analogy to
explain the dynamics of the laser pulse. According to this picture, the laser
pulse is represented by a bound state of a particle trapped in a
quantum well. The inner component of the laser pulse corresponds to
that part of the bound
state where the energy of the system is larger than the  bottom of the
potential well. The outer part is the analogue of the 
exponential-decaying wings of the bound state penetrating a smooth
potential wall. Here, the self-focusing process of the laser pulse is
associated to the  shrinkage of the width of the quantum well. Theory
and simulations show that  during the self-focusing process
the energy of the inner component flows to
the outer one which stays with the pulse and does not diffract. Finally, only a
fraction of the pulse energy self-focuses. 
The present theory provides better agreement with results of numerical
simulations than most of the theories used in the last decades for the
self-focusing process of laser pulses.
Comparison of the prediction
of the present theory with Marburger's formula for the
self-focusing distance shows an agreement
within an error of $20\%$ over a broad range of input powers of the pulse.
Further comparisons with numerical simulations show that the pulse
length as well as  the on-axis intensity are predicted correctly within a $20\%$
error too. We may emphasize that the comparisons are performed 
for collimated laser pulses, which is the worst case possible. Consideration of
an external focusing would give even lower errors. 
Finally, we note that the present theory
can be easily extended to consider losses
and ionization of the gas medium. 

\appendix

\section{Analysis of the diffraction term\label{app2}}

From Eqs. (\ref{ansatz1}),
(\ref{psi0gauss})  and (\ref{finals2}) we can write
\begin{eqnarray}\label{ansatz2app2}
&u(\xi,r)&=\frac{1}{a(\xi)}A(\xi)
\exp{\left(\rho^2\right)}\exp{\left(i\,b_0\,a(\xi)\,\rho^2\right)}
\nonumber \\
&\times&
\exp{\bigg(i\,c(\xi)a(\xi)(2\rho^2\exp{(-2\rho^2)}-\exp{(-2\rho^2)})\bigg)}
\nonumber\\
&\times&\exp{\bigg(i\frac{\kappa\xi}{a^3(\xi)}\left(1-2\rho^2\right)\bigg)},\quad\quad
\end{eqnarray}
where $A(\xi)$, $a(\xi)$ and $c(\xi)$ are assumed to be
variational parameters, and $\kappa$ is an unknown constant.
By doing
variational approximation similar to that one in section
\ref{refsec1}, but using the ansatz (\ref{ansatz2app2}), one can show that in
the critical regime $(P=P_{cr})$
\begin{equation}\label{diffracrestric}
\kappa=\textrm{constant}\times a^{\prime\prime}(0)=0.
\end{equation}
Thus, in principle we can neglect corrections on the diffraction term
and keep only the corrections in the nonlinear term in (\ref{ansatz2app2}).


\begin{thebibliography}{99}

\bibitem{Kasparian03} J. Kasparian, M. Rodriguez,
G. M{\'e}jean, J.. Yu, E. Salmon, H. Wille, R. Bourayou, S. Frey, Y.B
A. Mysyrowicz, R. Sauerbrey, J.P. Wolf, and L. W\"oste, Science
{\bf 301}, 61 (2003).

\bibitem{Alfano89} R. R. Alfano, Supercontinuum Laser Source (Springer Verlag, New York, 1989).

\bibitem{marburger75} J.H. Marburger, Prog. Quant. Electr. {\bf 4}, 35 (1975).

\bibitem{dawes69} E. Dawes and J. Marburger,
{\it Phys. Rev.} {\bf 179}, 862 (1969).

\bibitem{fibich00} G. Fibich and A. Gaeta, Opt. Lett. {\bf 25}, 335 (2000).

\bibitem{fibich96} G. Fibich, Opt. Lett. {\bf 21}, 1735 (1996).

\bibitem{fibich99} G. Fibich and G. Papanicolaou, SIAM J. Appl. Math. {\bf
60}, 183 (1999).

\bibitem{boyd} R. W. Boyd, Nonlinear Optics (Academic Press, San
Diego, USA, 2003).

\bibitem{akozbek00}  N. Ak\"ozbek, C. M. Bowden, A. Talebpour and S. L. Chin,
Phys. Rev.  E {\bf 61}, 4540 (2000).

\bibitem{Schwarz-Diels01} J. Schwarz and J. C. Diels,
Phys. Rev. A. {\bf 65}, 013806 (2001).

\bibitem{Sprangle} P. Sprangle, J. R. Pe\~nano and B. Hafizi,
Phys. Rev. E {\bf 66}, 046418 (2002).

\bibitem{Skupin04} S. Skupin, L. Berg\'e, U. Peschel, and F. Lederer,
Phys.Rev.Lett. {\bf 93}, 023901 (2004).

\bibitem{Mlejnek98} M. Mlejnek, E. M. Wright, and J. V. Moloney,
Opt. Lett., {\bf 23}, 382 (1998).

\bibitem{Kandidov03} V.P. Kandidov, O.G. Kosareva, and A.A. Koltun,
Quant. Electr. {\bf 33}, 69 (2003).

\bibitem{Liu} W. Liu, J. F. Gravel, F. Th\'eberge, A. Becker and
S. L. Chin, Appl. Phys. B (accepted for publication).





\end{thebibliography}
\end{document}